\pdfoutput=1

\documentclass[sigplan, screen]{acmart}

\renewcommand\footnotetextcopyrightpermission[1]{} 

\usepackage{includes/packages}
\usepackage{includes/commands}

\title{MCMComm: Hardware-Software Co-Optimization for End-to-End Communication in Multi-Chip-Modules} 

\author{Ritik Raj$^{\dagger*}$, Shengjie Lin$^{\dagger*}$, William Won$^\dagger$,  
    Tushar Krishna$^\dagger$}

\affiliation{
    \institution{$^\dagger$Georgia Institute of Technology, GA, USA}
    \country{}
}
\email{{rraj67, slin468, william.won}@gatech.edu, tushar@ece.gatech.edu}

\thanks{$^{*}$Both authors contributed equally to this research.}

\begin{document}

\begin{abstract}

Increasing AI computing demands and slowing transistor scaling have led to the advent of Multi-Chip-Module~(MCMs) based accelerators.
MCMs enable cost-effective scalability, higher yield, and modular reuse by partitioning large chips into smaller chiplets.
However, MCMs come at an increased communication cost, which requires critical analysis and optimization.
This paper makes three main contributions (i)~end-to-end, off-chip congestion-aware and packaging-adaptive analytical framework for detailed analysis, (ii) hardware software co-optimization incorporating diagonal links, on-chip redistribution, and non-uniform workload partitioning to optimize the framework, and (iii)~using metaheuristics~(genetic algorithms; GA) and mixed integer quadratic programming~(MIQP) to solve the optimized framework.
Experimental results demonstrate significant performance improvements for CNNs and Vision Transformers, showcasing up to 1.58$\times$ and 2.7$\times$ EdP~(Energy delay Product) improvement using GA and MIQP, respectively.

\end{abstract}

\maketitle
\pagestyle{plain}
\section{Introduction}
With the advent of artificial intelligence~(AI) applications in recent years, computing demands have grown tremendously~\cite{sevilla2022compute, openai_compute, desislavov2021compute}.
Research indicates that computational requirements for AI models have doubled every 3-4 months since 2012~\cite{openai_compute}.
These requirements are driven by the development of increasingly complex models and the expansion of AI applications across various sectors including natural language processing~(NLP)~\cite{dubey2024llama, team2023gemini, achiam2023gpt}, autonomous driving~\cite{zheng2025genad, hu2023planning, atakishiyev2024explainable} and healthcare~\cite{rahman2023ambiguous, zhang2022contrastive, gao2018human}, among others.
A lot of AI accelerators including Google TPU v5~\cite{tpuv5}, Meta MTIA~\cite{firoozshahian2023mtia}, Cerebras WSE-2~\cite{Cerebras}, Nvidia DGX GH200~\cite{nvidia-dgx} and Tesla Dojo~\cite{talpes2022dojo} have incorporated multi-chip and multi-core~\cite{emami2023manticore} designs to meet the increasing demands.

Recent advancements in multi-chip module~(MCM) integration have emerged as a promising solution to increasing compute demands in an era of slowing transistor scaling~\cite{frank2001device, razavieh2019challenges}.
MCMs enable the creation of large-scale CPUs~\cite{naffziger2021pioneering, kannan2015enabling, beck2018zeppelin} and GPUs~\cite{arunkumar2017mcm, zhang2023seechip, zhang2023balancing} by integrating multiple semiconductor dies onto a single substrate.
With the sharp increase in fabrication costs for a large monolithic die sub-16nm technology~\cite{amd-keynote}, there has been a large-scale adoption of chiplet-based systems~\cite{shao2019simba, dave2023chiplet, pal2021designing, li2020chiplet, mok2021chiplet, shan2022architecture} using MCMs that provide low-cost fabrication alternatives.

However, the tradeoff of MCMs over monolithic designs is the increased communication overhead~\cite{musavi2024communication, graening2023chiplets, loh2015interconnect}.
The communication overhead depends on a wide variety of factors including workload dimensions, partitioning, scheduling, chiplet count, interconnects, and dataflow which has been modeled, in some aspect, by previous frameworks including SET~\cite{cai2023inter}, SIMBA~\cite{shao2019simba} and SCAR~\cite{odema2024scar}.
Package-level integration of MCMs via 2.5D or 3D~\cite{mishty2024chiplet}, variations in main memory bandwidth and placement extend the vast communication space even further with increased complexity.
Off-chip communication occupies a significant portion of the total latency in AI accelerators~\cite{gholami2024ai, shalf2011exascale, kestor2013quantifying, musavi2024communication, dalmia2024cpelide}.
A recent paper~\cite{pao2025eda} shows that off-chip memory accounts for an average of 54\% of total NPU energy across a variety of deep neural network~(DNN) workloads using a layer-wise scheduler.
The placement and bandwidth variations also result in shifting of the congestion bottleneck (\autoref{subsection: Motivation-main memory congestion}) and require rethinking of MCM communication modeling.

\insertFigureWide{MCMComm_Introduction}{
MCMComm system with NPU-based chiplets, challenges, and key ideas, optimized LS scheduling space, and real-time applications.
}{1}{-2em}{-0em}

\begin{table*}[]
\centering
\caption{Related Works comparing MCMComm with SIMBA, Tangram, SET and SCAR works}
\vspace{-1em}
\resizebox{\textwidth}{!}{%
\begin{tabular}{|l|l|l|l|l|l|l|l|l|}
\hline
Work & Use Case & \begin{tabular}[c]{@{}l@{}}Workload\\ partitioning\end{tabular} & \begin{tabular}[c]{@{}l@{}}Optimization\\ Algorithm\end{tabular} & Hardware & Scheduling Space & \begin{tabular}[c]{@{}l@{}}Hardware\\ Co-Design\end{tabular} & \begin{tabular}[c]{@{}l@{}}Main memory BW\\ accurate modeling\end{tabular} & \begin{tabular}[c]{@{}l@{}}Packaging-Aware\\modeling\end{tabular} \\ \hline
SIMBA \cite{shao2019simba} & single model & non-unifrom & heuristics & MCM & LS/LP & No & No & No \\ \hline
Tangram \cite{gao2019tangram} & single-model & uniform & heuristics & Multi-Core & LP & No & No & No \\ \hline
SET \cite{cai2023inter} & single-model & uniform & metaheuristics & Multi-Core & LS/LP & No & No & No \\ \hline

SCAR \cite{odema2024scar} & multi-model & uniform & heuristics & MCM & LS/LP & No & No & No \\ \hline
\begin{tabular}[c]{@{}l@{}}\textbf{MCMComm} \\ \textbf{(This Work)}\end{tabular} & \textbf{single-model} & \textbf{non-uniform} & \textbf{MIQP/GA} & \textbf{MCM} & \begin{tabular}[c]{@{}l@{}}\textbf{Redistribution based} \\ \textbf{fine-grained pipelined LS}\end{tabular} & \begin{tabular}[c]{@{}l@{}}\textbf{Diagonal} \\ \textbf{(NoP Links)}\end{tabular} & \textbf{Yes} & \textbf{Yes} \\ \hline
\end{tabular}%
}
\label{tab:related works}
\end{table*}

There has been several prior work in optimizing inter-layer scheduling~\cite{cai2023inter,gao2019tangram,zheng2023tileflow,yang2023isosceles,garg2024pipeorgan,garg2022understanding}.
One such work~\cite{cai2023inter} defined and explored the inter-layer scheduling space for tiled accelerators including two broad schemes: layer sequential~(LS) and layer pipeline~(LP).
LS devotes every core/chiplet to one layer at a time.
The workload partitioning among all the cores/chiplets results in lower on-chip memory requirements
but extends the critical path of computation.
Additionally, LS may result in under-utilization arising out of significant skewness in layer dimensions.
In contrast, LP scheduling maps the entire DNN model onto the accelerator, allowing multiple layers to be processed concurrently in a pipelined manner.
This concurrency can significantly enhance throughput by overlapping the execution of different layers, thereby reducing end-to-end latency but at a cost of higher on-chip memory requirement as shown in~\autoref{fig:MCMComm_Introduction}.
MCMComm alleviates the inter-layer communication bottleneck in LS scheduling space by introducing on-package redistribution (\autoref{subsection: on package redistribution}) and fine-grained pipelining (\autoref{subsection: fine-grained pipeline}) at a cost of higher on-chip memory requirement.
MCMComm tackles data duplication overhead of LS space by proposing mixed integer quadratic Pprogramming~(MIQP)-optimized (\autoref{subsection:NonlinearProgramming}) row or column-wise non-uniform GEMM partitioning.
MCM optimizes LS scheduling space while being orthogonal to LP space due to our proposed cycle-accurate framework and fine-grained optimizations.

MCMComm addresses significant limitations observed in traditional optimization methods for multi-chip module (MCM) systems with applications in self-driving, autonomous systems, and chatbot inference as shown in~\autoref{fig:MCMComm_Introduction}.
To summarize, this work makes four main contributions:

\begin{itemize}[leftmargin=*]
    \item By moving beyond layer-by-layer and heuristics-based workload partitioning optimization \cite{shao2019simba}, MCMComm takes into account cross-layer implications of workload partitioning using end-to-end cycle-accurate modeling.
    \emph{It is the first framework, to the best of our knowledge, to include main memory congestion-aware and packaging-adaptive modeling covering a wide variety of MCMs}.
    
    \item Hardware-software co-optimization on LS scheduling space using diagonal links, on-chip redistribution, non-uniform workload partitioning, and fined-grained pipelining to optimize inter-layer communication bottleneck and data duplication as shown in~\autoref{fig:MCMComm_Introduction}.
    
    \item Scheduling using metaheuristic (Genetic Algorithm) and Mixed Integer Quadratic Programming (MIQP) to solve the optimized MCMComm framework. These two methods trade off scheduling time and optimal solutions.
    \item up to 35\% and 142\% geo-mean improvement using GA and MIQP respectively over non-optimized uniform partitioning (baseline) for latency. Moreover, up to 37\% and 72\% geomean improvements using GA and MIQP, respectively, over the baseline for EdP.
\end{itemize}

\section{Background and Related Works}

\insertFigureWide{MCMComm_framework}{
MCMComm framework with genetic algorithm and non-integer programming schedulers showing different input knobs. The framework is packaging-adaptive as shown by four types of chiplets showing different positions of main memory (DRAM/HBM) in 2.5D and 3D packaging. The framework separately models high-BW and low-BW off-chip cases making it congestion-aware.
}{1}{-2em}{-0.5em}

\subsection{Multi-Chip-Modules}
MCMs are advanced packaging solutions that integrate multiple integrated circuits~(ICs) or semiconductor dies into a single package. MCMs have emerged as a promising solution to address the increasing complexity and performance demands of modern integrated circuits. By integrating multiple semiconductor dies onto a single substrate, MCMs offer significant advantages over traditional monolithic chip designs. This approach addresses challenges related to scalability, performance, and manufacturing efficiency. For instance, NVIDIA's MCM-GPU \cite{arunkumar2017mcm} architecture demonstrates substantial performance improvements over traditional monolithic designs. Additionally, the development of multi-chip stacked memory modules using chip-to-wafer (C2W) \cite{sekhar2024multi} bonding techniques has shown promise in achieving high-density memory integration. 

\subsection{Scheduling in Multi-Core/MCM Accelerators}
Inter-layer scheduling is a critical aspect of optimizing deep neural network (DNN) accelerators, focusing on the allocation of computing resources and the execution order of DNN layers to enhance utilization and energy efficiency. Traditional approaches often rely on heuristic patterns, which may not fully exploit the potential of tiled accelerator architectures. According to SET \cite{cai2023inter}, there are broadly two types of inter-layer scheduling schemes: Layer-Sequential (LS) and Layer-Pipeline (LP). In specific, they introduce a Resource Allocation (RA) tree-based notation to effectively represent and analyze these scheduling schemes and find the near-optimal configuration of a set of LS/LP using a metaheuristic algorithm called Simulated Annealing (SA). Additionally, the HW-Flow-Fusion framework \cite{valpreda2022hw} has been proposed to fuse operations from the different convolutional neural network (CNN) layers, optimizing execution on dataflow architectures by maximizing data reuse and minimizing data movement. However, they don't consider latency profiles of different types of chiplet systems which can lead to lots of complex communication modeling as discussed in this paper. But they did not consider the latency profile of chiplet-based systems (\autoref{fig:MCMComm_framework}), which is different from single-chip accelerators. Communication delay is the shortest for chiplets near the main memory and is largest for far away chiplets necessitating non-uniform workload distribution.

There have been several works on LP space as well including Tangram \cite{gao2019tangram} and TileFLow \cite{zheng2023tileflow}. 
MCMComm is orthogonal to LP scheduling schemes. Given an LP scheme, MCMComm can optimize the workload partitions of different layers ensuring end-to-end performance improvement. For example, suppose a 4x4 MCM system is divided equally among two layers. We can model each 2x4 MCM system separately. The 2x4 system closer to the main memory can be modeled using type A and the other 2x4 MCM can be modeled using type B (\ref{fig:MCMComm_Introduction}) where the first 2x4 system will serve as the distributed interface of data transfer. 

While Simba \cite{shao2019simba} did introduce communication-aware non-uniform workload division, it used heuristics-based layer-by-layer optimization and failed to observe end-to-end implication (\autoref{Motivation-layerbylayer}). This work focuses on applications like self-driving and autonomous systems that require single model acceleration on edge MCMs as opposed to MAGMA \cite{kao2022magma} and SCAR \cite{odema2024scar} which focuses on multi-tenant scheduling in the cloud. \autoref{tab:related works} highlights the key difference between our work and related works.

\section{Motivation}

\subsection{Layer-by-Layer Workload Partition Optimization}
\label{Motivation-layerbylayer}
A recent study \cite{musavi2024communication} identified that inter-layer communication significantly dominates data movement in MCM accelerators, causing a significant bottleneck with increasing chiplet size.

Previous works on chiplets including SIMBA \cite{shao2019simba} optimize workload partitioning layer by layer using greedy heuristics. SIMBA partitions the workload non-uniformly and inversely proportional to the communication distance of a chiplet from the off-chip memory. Although this approach optimizes a single layer but does not consider inter-layer implications in end-to-end latency and results in sub-optimal design. For example, consider a compute-heavy workload where off-chip communication is only required at the beginning and end of the workload, and the rest of the time is devoted to computation. Also, assume that uniform partitions result in each chiplet being 100\% utilized. Any smaller partition will result in under-utilization. Following a SIMBA-like approach will result in farther chiplets getting smaller partitions of the workload and therefore, being under-utilized. Moreover, in the case of distributed off-chip memories (type B and D in Fig. \ref{fig:MCMComm_framework}), simply partitioning the workloads inversely proportional to the distance from the off-chip memories is not the optimal strategy and requires a deeper analysis of workload partitioning while considering the end-to-end implications.

\subsection{DRAM/HBM need different analytical models}
\label{subsection: Motivation-main memory congestion}

\insertFigure{motivation_heatmap}{
Modeling results when all chiplets are pulling 1~GB message from the memory over a 4$\times$4 Mesh.
Node 16 denotes the memory node.
DRAM/HBM bandwidth is 60~GB/s and 1,024~GB/s, and Low/High NoP link bandwidth is 60~GB/s and 120~GB/s, respectively. 
(a)--(c)~Network utilization heat map with different memory types and placements, when NoP bandwidth is 60~GB/s.
(d)~Total network communication latencies.
}{1}{-2em}{-1em}

To show the impact of using DRAM and HBM over a NoP system, we conducted simulations using the network modeling of ASTRA-sim~\cite{wonastrasim2, wontacos}.
All 16 chiplets concurrently pull 1~GB message from the memory. The results are depicted in~\autoref{fig:motivation_heatmap}.
As shown in~\autoref{fig:motivation_heatmap}(a), when the memory is DRAM, memory bandwidth is the bottleneck and determines the total transfer time.
This is further demonstrated in~\autoref{fig:motivation_heatmap}(d) that incrementing the NoP bandwidth by 2$\times$ yielded no performance benefit.
Meanwhile, \autoref{fig:motivation_heatmap}(b) represents that when the memory is HBM, the congestion effect moves to the package networks.
\autoref{fig:motivation_heatmap}(d) proves that the performance scales linearly by the NoP bandwidth.
\cite{musavi2024communication} shows that GoogleNet and DarkNet19 have siginificant off-chip communication overhead up to by 46.5\% and 44.5\%, respectively. 

\subsection{Packaging Needs Tailored Optimization}

We also compared how the placement of the memory module impacts the overall network performance.
\autoref{fig:motivation_heatmap}(c) summarizes the result.
Compared to the peripheral placement of HBM in~\autoref{fig:motivation_heatmap}(b), the central placement of HBM effectively mitigated the congestion over the NoP network.
As shown in~\autoref{fig:motivation_heatmap}(d), this resulted in a 1.53$\times$ improvement over the peripheral placement of the HBM.
For DRAM, due to the relatively low bandwidth, the memory is the bottleneck, thereby the placement did not have much impact.
This observation suggests that the judicious considerations in the placement of memory modules, as well as their types, are pivotal to optimizing NoP-based platforms.

\subsection{Inefficient Software-Only Optimization}

As per Amdahl's law \cite{amdahl1967validity}, there is a limited speedup that can be achieved via hardware-only optimization. This necessitates a software co-design using effective scheduling to alleviate the communication bottleneck. Optimizing only hardware or software shifts the bottleneck to the other and limits performance gains. There have been a lot of works incorporating hardware-software co-design strategies \cite{de1997hardware, wolf2002hardware} to optimize the end-to-end performance.

In MCM scheduling, suppose we have an ideal scheduling strategy for a type A system (\autoref{fig:MCMComm_framework}) that minimizes communication. Also, suppose that the majority of the communication includes writing the data from chiplets to the main memory. In this case, the two Network on Package (NoP) links connecting the global chiplet (access to main memory) to other chiplets become the bottleneck. 

\subsection{Heuristics-Based Suboptimal Solution}

The design space of the MCM system is large and complex. Heuristics \cite{lenat1982nature, gigerenzer2008heuristics, michalewicz2013solve} and metaheuristics \cite{abdel2018metaheuristic, hussain2019metaheuristic, yang2010engineering, yang2010nature} based approaches are well suited for large design space problems. Metaheuristics are based on optimizing a set of configurations through multiple iterations. Metaheuristics-based algorithms including simulated annealing \cite{van1987simulated, bertsimas1993simulated}, genetic algorithm \cite{mitchell1998introduction,holland1992genetic, mirjalili2019genetic}, etc. have been used by previous works \cite{cai2023inter, kao2022magma} for optimizing inter-layer communication bottleneck in multi-core or MCM systems. But they are they result in sub-optimal solutions for a complex optimization space as we see later in \ref{evaluation}. In addition, a simple greedy-based heuristic performs even worse. We incorporate an MIQP-based approach that outperforms the metaheuristics~(GA)-based approach by up to 1.63$\times$ and a heuristics-based approach by up to 3.25$\times$. The trade-off comes in solving time, where heuristics is instantaneous, GA takes around 30s and MIQP takes around 4 minutes on average.

\section{End-to-end Analytical Modeling}

This section is divided into four sections. \autoref{subsection: four chiplet systems} talks about four different types of Chiplet systems covering a wide range of MCM systems. \autoref{Scheduling Space Formulation} models the high-level end-to-end scheduling of machine learning workloads on MCMs with given objectives. \autoref{Latency Modeling} and \autoref{Energy-delay Product Modeling} deep dives and model latency and edp respectively.

\subsection{Types of MCM Systems}
\label{subsection: four chiplet systems}

\autoref{fig:MCMComm_framework} shows latency profiles of four types of MCM systems based on relative positions of main memory (HBM/DRAM) and chiplets. Chiplets far away from the main memory will incur higher communication overhead than the closer chiplets. SIMBA \cite{shao2019simba} and Manticore \cite{emami2023manticore} are examples of type A systems where the main memory is placed in a corner away from cores/chiplets. MTIA \cite{firoozshahian2023mtia} is an example of a type B system where main memory is evenly distributed outside the 2D array of cores. Type A and type B systems are based on 2.5D packaging where memory and logic are stacked on top of an interposer. In comparison, the type C system is based on 3D packaging where memory is stacked on top of logic \cite{kim2011stacked, chen20203d}. Type D system is a combination of type B (2.5D) and type C (3D) systems and Chiplet-Gym \cite{mishty2024chiplet} discusses the design space exploration (DSE) of such systems.

\subsection{Scheduling Space Formulation}
\label{Scheduling Space Formulation}

\subsubsection{Hardware Configurations\label{subsubsection:HardwareConfig}}
We first define the configuration parameters that capture the characteristics of MCM systems as follows:
    \[HW=\{BW_{nop}, BW_{mem},X,Y,R,C,type\}\]
where $BW_{nop}$ and $BW_{mem}$ represents network-on-package bandwidth among chiplets and offchip bandwidth between global chiplet(s) and main memory respectively. $X$ and $Y$ represent the number of chiplets in the x and y direction, respectively, in an MCM grid of chiplets. $R$ and $C$ represent rows and columns of a systolic array in a chiplet. We assume that each chiplet contains one systolic array. $type$ specifies the way of communication between off-chip memory and chiplets.

Given the parameters of the hardware, we further index each chiplet to be $(x,y)$. As shown in figure \ref{fig:Diags}, $x$ and $y$ represents the number of rows and columns away from the global chiplet(s) connected with the main memory. We assume that each chiplet will only communicate with the closest global chiplet. 

Such indexing encodes all necessary topological information for scheduling. Therefore, we can adapt to different types of systems with different access to main memory by using corresponding specialized indexing, as shown for four types of 5x5 systems in figure \ref{fig:Diags}.

\subsubsection{Machine Learning Task\label{subsubsection:MLTask}}
Machine learning workload can be represented as a directed acyclic graph~(DAG). We can execute any DAG by one of its topological orders. Therefore, we define a given machine learning workload as a sequence of operators:
\begin{equation}
    Task=[OP_0,OP_1,\dots,OP_{N-1}]
\end{equation}
Because machine learning workloads are dominated by general matrix multiplications (GEMM), we focus on executing a sequence of GEMM spatially partitioned among chiplets in an LS scheduling space. We support operators such as \textit{RELU} computed in the SIMD unit of the chiplet. We also support another set of operators including \textit{softmax} and \textit{layer norm} that introduce synchronization of chiplets for outputs distributed among them. We need to capture this synchronization during the execution of GEMMs. To achieve this, we define the attributes of a GEMM operator as 
\begin{equation}
    OP_i=\{M,K,N,sync,shared\_row,shared\_col\}
\end{equation}
where $M,K,N$ represents the input dimension, hidden dimension, and output dimension of a GEMM. $sync$ is a boolean value representing whether the output of $OP_i$ needs to be synchronized among chiplets. $shared\_row$ is $true$ when chiplets of the same row produce the same rows in the output matrix, similar to $shared\_col$.

\subsubsection{Workload Allocation\label{subsubsection:WorkloadAlloc}} Given hardware configuration $HW$ and machine learning task $Task$, we need to distribute workload among chiplets. 

Following LS scheduling space and spatial partitioning, we define the workload partition of $OP_i$ as $Px_i[X]$ and $Py_i[Y]$, where $Px_i[x],x\in X$ represents the numbers of rows in the output matrix to be processed by $x$-th row of chiplets. Similarly, $Py_i[y],y\in Y$ denotes the column workload partition of the output matrix among columns of chiplets.

To ensure that the distributed workload sums to the original GEMM, we constrain that
\[\sum_{x=0}^{R-1}Px_i[x]=M_i,\sum_{y=0}^{C-1}Py_i[y]=N_i\]
where $M_i$ and $N_i$ represent $M$ and $N$ of GEMM $OP_i$.

\subsubsection{Scheduling Problem\label{subsubsection:Scheduling}} Given configurations in \autoref{subsubsection:HardwareConfig}, workload definition in \autoref{subsubsection:MLTask} and allocation in \autoref{subsubsection:WorkloadAlloc}, we can now modeling the end-to-end cost as follows:
\begin{align}
    Cost&=Sche\left(\{
    comp(*_i),comm(*_i)|i\in[N]\}\right)\\
    comp(*_i)&=Combine_{comp}\left(comp_{x,y}(*_i)\right)\\
    comm(*_i)&=Combine_{comm}\left(comm_{x,y,t}(*_i)\right)
\end{align}
Here $(*_i)=(HW,OP_i,Px_i,Py_i)$ represent the configuration and scheduling of $i$-th GEMM operator. $comp$, $comm$ denote the cost function of GEMM computation and data communication. In the LS scheduling space, each chiplet takes and outputs a chunk of data. $comp_{x,y}$ calculates the individual compute cost of chiplet indexed $(x,y)$. Similarly, $comm_{x,y,t}$ calculates the individual cost of communication with a specific type $t$. Then, $Combine_{comp}$ and $Combine_{comm}$ take the costs of chiplets, resolve potential resource contention, and get the overall step cost. Finally, $Sche$ adjusts executing orders without breaking data dependencies, resulting in better resource utilization.

$Combine$ functions are inherited by the properties of the objective, which cannot be changed once set. Therefore, we formulate the scheduling problem as
\begin{equation}
    \arg\min_{Px,Py,Sche}Cost|HW,OP
\end{equation}
By using different cost functions, we are able to minimize various metrics including latency and energy-delay product (EDP).

\insertFigure{Diags}{
Illustration of chiplet topology for 4 types of systems. The dashed lines represent chiplet allocation results. Each chiplet has a local index (x, y).
}{0.9}{-1em}{-1em}

\subsection{Latency Modeling\label{Latency Modeling}} This section demonstrates the latency modeling of machine learning tasks on MCMs. Specifically, we will model the cost function $comp$ and $comm$ in \autoref{subsubsection:Scheduling} for the objective of latency.

\subsubsection{Computation} Compute is modeled by the output stationary dataflow \cite{sim2019energy, chen2017using}. Following cycle-accurate equations described in SCALE-Sim \cite{samajdar2020systematic, raj2025scale}, the latency cost for computation in a chiplet at index $(x,y)$ is 
\begin{equation}
    comp_{x,y}(*_i)=(2*R+C+K-2)*(Px_i[x]/R)*(Py_i[y]/C)
\end{equation}

\insertFigure{Congestion}{\label{fig:Congestion}Illustration of Congestion during data collection and effects of diagonal links.
}{0.8}{-1em}{-1em}

\subsubsection{Data Offloading\label{Data Offloading}}
To model the data offloading process, we complete the offloading in two steps. First, we collect output data from all chiplets. Then, the data is sent to off-chip memory.

\noindent \textbf{Step 1: On-chip Data Collection.} Consider MCM type A system where the global chiplet is the only link between other chiplets and main memory. In this case, we make a key observation that the amount of workload assigned to each chiplet will monolithically non-increase with the increasing distance. It is also mentioned in SIMBA \cite{shao2019simba} where the workload allocation is greedily chosen to be inversely proportional to the distance. Therefore, when sending data to the global chiplet, the bottleneck of latency lies on the NoP links connecting the global chiplet to other chiplets. Therefore, for this step, the total latency needed for data collection will be
\begin{align}
    \nonumber Combine_{comm}&(comm_{x,y,out}(*_i))=\\
    &\frac{M_i*N_i}{\text{bandwidth to entrances}*BW_{nop}}
\end{align} \label{noshared}
where $M_i*N_i$ is the size of output data of $OP_i$.

\noindent \textbf{Step 2: Data Transfer To Off-chip Memory.} This is a simple case, we simply divide data volume by bandwidth:
\[comm_{off-chip}=sizeof(\textbf{data})/BW_{mem}\]

\subsubsection{Data Loading\label{Data Loading}} To model the data loading process, congestion needs to be considered because of limited non-uniform access to memory in MCMs. As a result, NoP links in different places of the topology have various popularity. Popular links may experience congestion due to over-subscription. For such cases, nontrivial communication strategies are needed for congestion resolution. To manage the complexity of interactions between communication strategy and workload partitioning, we adopt a fixed communication strategy. 

We can decompose any off-chip communication in two steps. First, we send input data from off-chip memory to global chiplet(s). Second, we need to distribute input data to destination chiplets. We will discuss the two steps separately.

\noindent \textbf{Step 1: Data Transfer From Off-chip Memory.} This step is the inverse operation of Step 2 in data offloading and follows the same modeling.

\noindent \textbf{Step 2: On-chip Data Distribution.} For on-chip data distribution or collection, the data needs to be distributed from global chiplet(s) to other chiplets. With different hardware configurations, congestion will take in different places. We do congestion-aware modeling for low and high off-chip bandwidth cases.

\textit{Case 1: Low Bandwidth Case (DRAM).}
\label{subsubsection:low bandwidth case}
When the off-chip bandwidth is lower than the NoP bandwidth, main memory to chiplet communication becomes a bottleneck. Therefore, there is little to no contention while modeling chiplet-to-chiplet communication. The equation for inter-layer communication for input and weight is given by 
\begin{equation}
    sizeof(\text{data})/BW_{nop} * \text{no. of hops} 
\end{equation}
For hardware with low off-chip memory bandwidth, the communication bottleneck happens on the off-chip data transfer. Therefore, when the data of a chiplet arrives at the global chiplet, the closest links will all be available as previous data has already finished transfer. Therefore, for chiplet $(x,y)$
\begin{equation}
    \text{no. of hops}=x+y
\end{equation}
as this is the minimal number of hops required for the data to be at the destination.

\textit{Case 2: High Bandwidth Case (HBM).}
However, when we connect off-chip memory and global chiplet(s) with high bandwidth links, the communication bottleneck transfers to on-chip data distribution. In this case, we need to discuss two subcases separately, based on the data utilization characteristics.

\label{case2.1}\textit{Case 2.1: Row-wise or column-wise shared data.} We first send the input data to its target row if the data is column-wise shared (similarly to row-wise shared). Once the data gets to the target row/column, it then broadcasts to all the chiplets in the same row/column. In such cases, the former step will bring congestion. For example, as shown in figure \ref{fig:Diags}, for data that are row-wise shared, they all need to be sent to the target column first. As a result, the fisrt column is congested. In such a case, we resolve the congestion by sending the data for the farthest row first, then the second farthest row, and so on. The idea is to alleviate non-uniform latency for off-chip data transfer. Therefore, for chiplet $(x,y)$
\begin{align}
    \nonumber\text{no. of hops}&=\text{waiting hops} +(x+y)\\
    &=(X-x)+(x+y)=X+y
\end{align}
Likewise, for column-shared data,
\begin{align}
    \nonumber\text{no. of hops}&=\text{waiting hops} +(y+x)\\
    &=(Y-y)+(y+x)=Y+x
\end{align}

\textit{Case 2.2: Non-shared data transfer.} We can view this case as the inverse of on-chip data collection (\autoref{noshared}), with the same cost function.

\subsection{Energy-delay Product Modeling\label{Energy-delay Product Modeling}}
To model the energy-delay product (EDP), we first model end-to-end energy consumption and then multiply with latency modeled in \autoref{Latency Modeling}.

\subsubsection{Computation}
For any chiplet of given workload $(input,filter,output)$. The energy produced is
\begin{align*}
    comp=&c_{SRAM}*(sizeof(inp+filt+out))+\\
    &c_{MAC}*cycles*R*C*(X*Y)
\end{align*}

\subsubsection{Off-chip Data Transfer}
For data transfer between off-chip memory and multi-chip-modules, energy is calculated as
\[comm_{off-chip}=c_{off-chip}*sizeof(\text{data})\]

\subsubsection{On-chip Data Transfer}
For on-chip data transfer and multi-chip-modules, energy is calculated as

\[comm_{on-chip}=c_{NoP}*sizeof(\text{data}) * \text{no. of hops}\]
where no. of hops is calculated following the same way as mentioned in \autoref{Latency Modeling}.

Here, $c_{SRAM}$, c$_{MAC}$, $c_{NoP}$ and $c_{off-chip}$ refer to energy per bit read/write from SRAM, energy per MAC unit per cycle, energy for NoP per bit per hop and energy per bit read/write from main memory respectively.

\section{Software-Hardware Co-optimization}
This section presents two novel techniques - Diagonal Links (\autoref{DiagonalLinks}) and On-Package Redistribution (\autoref{subsection: on package redistribution}) targeted at reducing non-uniform NoP communication, given their significant importance in MCM systems. In addition, the overlap between computation and communication is maximized through find-grained pipelining.

\subsection{Diagonal Links}
\label{DiagonalLinks}
MCMs have non-uniform access to main memory. A link is used by a chiplet if it lies on the transfer path of data that is needed by that chiplet. In an ideal setting, where each link can start transferring without considering data availability, assuming the same NoP bandwidth, links that are used by more chiplets will result in a communication bottleneck.

One example is data offloading, as discussed in \autoref{Data Offloading} and figure \ref{fig:Congestion}, where the links connected to the global chiplet(s) become the bottleneck for communication efficiency, no matter what communication strategy is used.

Therefore, to alleviate the congestion problem, we propose adding diagonal links to the systems, shown as blues links in figure \autoref{fig:Diags} and figure \autoref{fig:Congestion}. For the case where we gather outputs to main memory because the process is slowed down by the bottleneck efficiency, it brings 50\% more bandwidth on the bottleneck communication. Diagonal links also help in reducing contention for the case where the inputs are dispatched. By allowing data far away from the main memory to transfer across the diagonal links, we both reduce contention on the first column of links and alleviate the degree of non-uniform latency to the main memory. Therefore, the overall efficiency increases. 

\subsubsection{Performance Improvement}

First, we analyze the dispatching of row-wise or column-wise shared data, as shown in \autoref{fig:Diags}. In addition to the strategy discussed in \autoref{Data Loading}, diagonal links provide an alternative strategy. First, we utilize the diagonal links and then use the horizontal or vertical links to get the data to the destination. For the chiplet $(3,2)$ in type A system, shown in figure \ref{fig:Diags}, first, it goes along the blue diagonal links. Like in the last strategy, it needs to wait for data of rows below to be sent, that is $X-x$ hops. Then, it goes for $\min(x,y)$ links along the diagonal links. Finally, it goes along $\text{abs}(x-y)$ links to get to the destination. So, the total number of hops needed is
\[(X-x)+\min(x,y)+\text{abs}(x-y)=X-x+\max(x,y)\]

It should be noted that these two strategies don't conflict with each other on normal links. For example, for data that is row-wise shared, data in the previous strategy will only use the first column of vertical links while the strategy that uses diagonal links will only use the rest of the vertical links. Because the data is row-wise shared, there will be no conflict on horizontal links. Therefore, we can always take the minimum hops out of two strategies.

\subsection{On-package Redistribution}
\label{subsection: on package redistribution}
As far as neural network execution is concerned, it is often the case that the output of the last operation is the input of the next one. As a result, outputs are transferred back to the main memory or last level on-package memory before they are sent back, only in a slightly different arrangement. This introduces unnecessary communication between chiplets and main memory or last level on-package memory. 

We propose on-package redistribution, which greatly reduces data movements needed for output redistribution in a GEMM. It should be noted that the optimal communication strategy requires consideration of both the data size of each chiplet and the transfer order of each link, which makes it impractical to optimize for the dynamic data size across different operators. 

\begin{figure}
    \centering
    \begin{subfigure}{0.25\linewidth}
        \centering
        \includegraphics[width=0.99\linewidth]{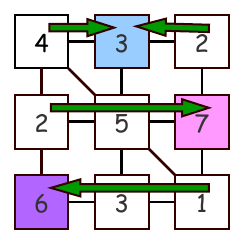}
        \caption{S1: Gather}
        \label{fig:step_1}
    \end{subfigure}
    \begin{subfigure}{0.25\linewidth}
        \centering
        \includegraphics[width=0.99\linewidth]{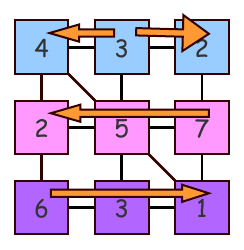}
        \caption{S2: Broadcast}
        \label{fig:step_2}
    \end{subfigure}
    \begin{subfigure}{0.25\linewidth}
        \centering
        \includegraphics[width=0.99\linewidth]{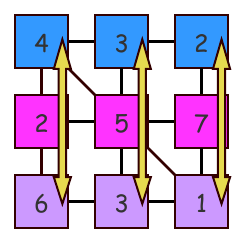}
        \caption{S3: Redistribute}
        \label{fig:step_3}
    \end{subfigure}
    \vspace{-1em}
    \caption{Three-step on-package data redistribution process. Arrows of different colors represent different communication steps. }
    \label{fig:redist}
    \vspace{-1em}
\end{figure}

We propose a simple three-step heuristic-based strategy for on-package data redistribution. As shown in figure \ref{fig:redist}, \underline{First}, we do row reduction: chiplets of the same row send data to the chiplet that best balances the left-coming data size and right-coming size, such that the latency for reduction is minimized. \underline{Second}, the reduced data is broadcasted to the other chiplets in the same row. \underline{Third}, because $P_r$ partition of one operator will be different from the other, we do the column redistribution: chiplets of the same column send data according to the placement requirement of the next operator.

The three-step strategy is designed with the assumption that vertical links help little during row reduction. This is because each data need to take two more hops to get to the same destination and contention may be introduced.

\subsection{Asynchronized Execution}
\label{subsection: fine-grained pipeline}

Although we decouple computation and communication as different operators, they can be fused into one operator. The benefit is that computation of each chiplet can be launched as soon as the data required are ready. This avoids redundant synchronization before computation. Analytically, this means simply adding equations of corresponding operations before synchronizations.

\insertFigure{pipeline}{
Illustration of pipelining. Blue and red blocks represent executions of two different samples. $Inp$, $Wi$, and $Outi$ stand for communication of corresponding input, filter, and output data. 
}{1}{-2em}{-1em}

\subsection{Pipelining\label{Pipelining}} For machine learning workloads, especially inference, many operators will be sequentially dependent and we can only execute them one by one. Therefore, many communication or computations during execution have no available complement operator to make use of idle resources.

However, in real tasks, we often process a batch of data instead of only one sample at a time. There are no data dependencies among different samples, which provides natural overlapping opportunities.

For example, in \autoref{fig:pipeline}, the naive sequential execution above will make no use of such overlapping opportunities. On the other hand, if we carefully arrange the execution order, such overlapping can be maximumly utilized.

As the subfigure below in \autoref{fig:pipeline}, we can utilize idle links during computation for communications of other batches. This is a relatively simple example, if we consider larger batch sizes with various execution durations, manual arrangement through heuristics may be suboptimal. 

We utilize a solver to arrange the order for maximum overlap. As shown in \cite{concerto}, this problem can be seen as a classic resource-constrained project scheduling problem (RCPSP), where compute and communication can be viewed as two different resources and each communication or computation can take only one resource to compute.

Although RCPSP is an NP-hard problem, in our case where only a sequence of GEMMs is considered, the number of executions is relatively small and can be efficiently handled by the solver.
\section{Analytical equations solving approach}

\subsection{Optimization Space}
In MCM systems, it is difficult to solve the optimal solution of workload partition for deep learning tasks, as it involves interactions of multiple factors. \underline{First}, interactions between multiple operators force the number of variables to increase with the number of operators. \underline{Second}, for any single communication task, an optimal communication strategy should be solved but this is itself a hard problem. \underline{Third}, the interaction between communication strategy and workload partition makes the problem even more difficult to solve.

Therefore, a careful trade-off needs to be made between the optimization space to be explored and the solving time that is allowed. The challenge is to decide which parts of the process are fixed. The fixed communication strategy can still be effective if the strategy is already adaptive to different workload partitions. In contrast, a fixed workload may leave little room for communication strategy as it is hard to get the best strategy behind specific workload partition. 

In this work, we define the optimization space to be the global workload partition with adaptive communication strategies. Each analytical equation that presents time spent on one specific operator assumes one fixed execution strategy. Empirical studies show that such optimization space catches key optimizations while allowing solving time to be within hours.

\subsection{Genetic Algorithm}
\label{subsection:GA}
Genetic Algorithm is an optimization method inspired by natural selection that evolves candidate solutions through selection, crossover, and mutation. It evaluates each candidate with a fitness function, iteratively refining solutions until a termination criterion is met, making it effective for complex problems like those in machine learning and engineering design. 

Crossover and mutation are performed on two sets of variables: the workload partitions and the positions of the collection chiplet during on-chip redistribution. Since the search space of the workload partition is so huge, the input partitions are constrained between $\min\left(R*(\left\lceil \frac{P_x}{R} \right\rceil - 2), R\right)$ and $\max(R*(\left\lceil \frac{P_x}{R} \right\rceil + 2), R)$ while ensuring the sum of all $P_x$ is equal to M, where Px refers to partitioning in chiplet row dimension, R is NPU row dimension and M is workload dimension in a GEMM operation. The minimum Px is equal to R since a smaller $P_x$ will lead to under-utilization in the systolic array. Similar constraints are applied to filter partitions as well. Even after highly constraining the partitions, the search space is close to \textit{O($2^{32}$)} for a 5x5 chiplet system and a 10-layer workload out of which on-chip redistribution is performed on two layers. The space increases quadratically for a larger chiplet system and linearly for a workload with larger layers.

\subsection{Mixed Integer Quadratic Programming}
\label{subsection:NonlinearProgramming}

Quadratic programming (QP) is a branch of mathematical optimization that focuses on problems involving quadratic functions. This approach is crucial for accurately modeling and solving our scenarios where relationships between variables are inherently nonlinear but mostly quadratic. Specifically, we use mixed-integer quadratic programming (MIQP).

\subsubsection{From Analytical Equation To MIQP}
To make analytical equations fit into the MIQP optimizer framework, some adjustments need to be made as divisions in the equations will make the solving process unacceptably slow. 

\underline{First}, for division by constants, we multiply the product of all such constants by all equations. Therefore, each constant denominator will be canceled out by this product. However, such a method expands equation values and the result to the times of the product. For that, some variables in the integer programming model exceed the presenting scope of integers, which slows down the solving process. We fixed this problem by shrinking all equation values by a constant scaling factor. This might bring some precision error to the model for the shrinking process. However, because the product is normally very large, a properly chosen factor has negligible impact on the final result.

\underline{Second}, for division by variables, we adopt a simple approximation replacement to change variable denominators to numerators:
\[\frac{\text{some equations}}{c+x}\sim \frac{\text{some equations}}{c^2}(c-x)\]
Here $c$ denotes some constants and $x$ is the variable denominator. It should be noted that such an approximation is effective only if x is close to c. In this work, divisions by variable denominators only happen by dividing the workload by hardware parameters. In such a case, it is reasonable to assume that hardware irregularity can only happen to a small degree.

\subsubsection{Modeling Framework}

Here is the pseudo-code for our nonlinear integer programming formation, as shown in Algorithm \ref{alg:ilp}. It should be noted that we encode different communication and computation strategies in $op$ which maps workload partition to expected execution time given fixed strategy. 

\begin{algorithm}[t!]
\caption{MIQP Formation}\label{alg:ilp}
\begin{algorithmic}
\Require $ops$, $row\_sizes$, $col\_sizes$ \Comment{model and input sizes}
\Ensure Solving $Px_is,Py_is$ for best end-to-end performance
\State $Px_is$, $Py_is$ $\gets$ $\arg\min_{Px_is,Py_is}$ ( \Comment{Workload Partitions}

sum$([$

$op(Px_i, Py_i)$ \Comment{output step time given workload}

for $(Px_i,Py_i,op,row\_size,col\_size)$ in $\backslash$

\quad$(Px_is,Py_is,ops,row\_sizes,col\_sizes)$ $\backslash$

\quad s.t. sum$(Px_i)=row\_size$ \& sum$(Py_i)=col\_size$.

])
\end{algorithmic}
\end{algorithm}

It should be noted that synchronization operators ($\max$) are added to each pair of computation and corresponding input communication. Although this synchronization potentially brings a negative impact on the overall performance, it is necessary for the sake of communication strategies in chapter \ref{DiagonalLinks}.

\section{Evaluation}
\label{evaluation}

In this section, we present an evaluation of MCMComm's end-to-end performance on MCMs targeted for image processing workloads such as Alexnet, vision workloads such as Vision Transformer and Vision Mamba, and autonomous workloads such as Hydranets (used in Tesla's self-driving cars) with different batch sizes. 

We evaluate MCMComm on a variety of systems. We evaluate 4x4, 8x8, 16x16 topology of chiplets. Each chiplet has a 16x16 systolic array. We assume NoP links cannot be shared by two data transfers at the same time. For each of these topologies, four types of chiplet systems with different off-chip bandwidths (HBM/DRAM) are tested. We set the common system configurations as given in \autoref{tab:system config}.

\begin{small}
\begin{table}[t]
\centering
\caption{MCMComm System Configurations}
\vspace{-1em}
\begin{tabular}{cc}
\toprule
High Memory BW (HBM) & 1000 GB/s \\
\hline
Low Memory BW (DRAM) & 60 GB/s \\
\hline
NoP Bandwidth & 60 GB/s \\
\hline
Chiplet Topology & 4x4, 8x8, 16x16\\
\hline
Systolic array size & 16x16\\
\hline
NoP Energy & 1.285 pJ/bit/hop\\
\hline
DRAM Energy & 14.8 pJ/bit \\
\hline
HBM Energy & 4.11 pJ/bit \\
\hline
SRAM Energy & 0.28 pJ/bit \\
\hline
MAC Energy & 4.6 pJ/cycle \\ \bottomrule
\end{tabular}
\label{tab:system config}
\vspace{-1em}
\end{table}
\end{small}

\begin{table}[t]
\centering
\caption{Evaluation Methodology}
\vspace{-1em}
\resizebox{0.9\columnwidth}{!}{
\begin{tabular}{ccc}
\toprule
\textbf{Scheduling} & \textbf{Workload} & \textbf{MCMComm}
\\ \midrule
Scheme & Partitioning & Optimizations \\
\hline
Layer Sequential & Uniform & No \\
(Baseline)  & & \\
\hline
SIMBA-like & Inversely Proportioanal  & No \\
& to Distance & \\
\hline
MCMCOMM-GA & GA optimized & Yes \\
\hline
MCMCOMM-MIQP & MIQP optimized & Yes \\ \bottomrule
\end{tabular}
}
\label{tab:eval method}
\vspace{-1em}
\end{table}

We choose Layer Sequential as the baseline with uniform workload partitioning and no optimizations. We also use SIMBA-like system to show the comparison with heuristic-based workload partition as summarized in \autoref{tab:eval method}

For MIQP for workload partitioning and ILP for pipeline scheduling, we limit the solving time to 10 minutes. We estimate the duration for each computation or communication step in pipelining on the basis of workload partitioning. We use end-to-end latency and EDP as optimization targets. 

\insertFigureWide{latency_4x4}{
Latency comparison of MIQP and GA over the baseline for High Bandwidth (HBM) case. Results are normalized.
}{1}{-2.5em}{-0em}

\subsection{End-to-end Latency Results With High-Bandwidth Memory}
This section presents results for 4x4 chiplets on four types of systems, provided in \autoref{fig:latency_4x4}. We show that the GA and MIQP approach outperforms LS for all types of systems by a geometric mean of 13\%/45\%, 5\%/15\%,9\%/43\%, and 19\%/25\%, respectively. Results show that the SIMBA-like heuristic achieves even slightly worse performance against LS, which demonstrates it cannot optimize our scenario where the target is the end-to-end implication of workload partitioning.

We noticed that in all settings, MCMComm provides the largest speedup on Alexnet. This is because on-chip data redistribution works for GEMMs that are sequentially chained. Alexnet has the most sequential structure where every operator takes only output from the previous convolution layer and static filter weight as inputs. Therefore, on-chip redistribution between every neighboring operator greatly reduces the amount of data transfer between operators resulting in lower end-to-end latency. For general attention operators in transformer-based models or vision mamba which utilized linear attention, the existence of attention heads makes the matrix multiplication a grouped GEMM operator, resulting in more complex data mapping. Therefore, such models only benefit from on-chip data redistribution in MLP layers.

We can also observe that for most cases, MIQP greatly outperforms other methods including GA. This shows the large optimization space of workload partition with diagonal links. Such space cannot be efficiently explored by heuristics like genetic algorithms. We also observed the closest performance between GA and MIQP in Type-D systems. We attribute this to the fact that communication latency to main memory is almost uniform in a 4x4 type-D system, in such case, the optimal workload partition will be closed to the uniform partition. Therefore, heuristics like genetic algorithms can find near-optimal solutions.

\insertFigure{latency_scaling}{
Latency comparison of MIQP and GA over the baseline for High Bandwidth case, type-A system. Results are normalized.
}{1}{-2.5em}{-1em}

\subsection{Scaling Results}
This section presents the performance of MCMComm on a type A system with different chiplet topologies. We show the performance both in latency and EDP, as in \autoref{fig:latency_scaling} and \autoref{fig:edp_scaling}. MIQP achieves a geometric mean of 55.5\% and 60.3\% speedup against LS while GA achieves 24.2\% and 35.1\%, respectively.

MIQP achieves similar speedup on the same model with different chiplet system scales except for Alexnet, which achieves higher speedup in larger chiplet systems. This is because the on-chip redistribution that benefits Alexnet the most saves both latency and energy by eliminating redundant communication. In addition, the saving is more pronounced with the increase of chiplet system scales.

It should be noted that in contrast, GA achieves relatively better performance in EDP experiments than Latency ones. This is because when the objective is EDP, the optimization potential is greater as both latency and energy can be reduced by workload partitioning. However, because that product of two end-to-end metrics (latency and energy) significantly complicates the modeling, the solver for MIQP finds it hard to get a solution close to optimal ones in a time limit of 10 minutes. Therefore, the solution is not fully optimized, relatively smaller gap from GA. Nevertheless, it still largely outperforms LS and Simba.

\insertFigure{edp_scaling}{
EDP comparison of MIQP and GA over the baseline for High Bandwidth case, type-A system. Results are normalized.
}{1}{-2em}{-1em}

\begin{figure}[!t]
    \centering
    \includegraphics[width=0.99\linewidth]{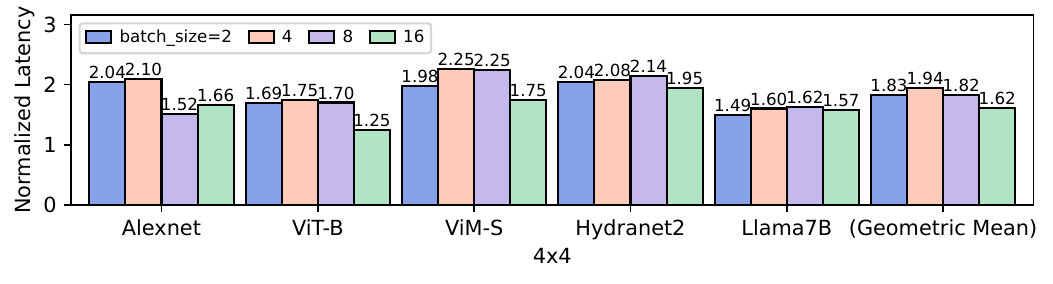}
    \vspace{-1em}
    \caption{Performance of pipelining given different batch sizes.}
    \label{eval:pipeline}
    \vspace{-1em}
\end{figure}

\subsection{Results For Pipelining}
This section presents the effectiveness of pipelining. \autoref{eval:pipeline} shows the per-sample speedup compared with LS. The speedup remains about the same with different batch sizes, demonstrating the scaling performance of our pipelining approach. Because there is no data dependency between operators, such pipelining always finds ample opportunities for overlapping. It takes from a few seconds to minutes for the ILP solver to find the optimal schedule.

\subsection{Results For Low-bandwidth Memory}
\begin{figure}[!t]
    \centering
    \begin{subfigure}{0.99\linewidth}
    \centering
    \includegraphics[width=0.99\linewidth]{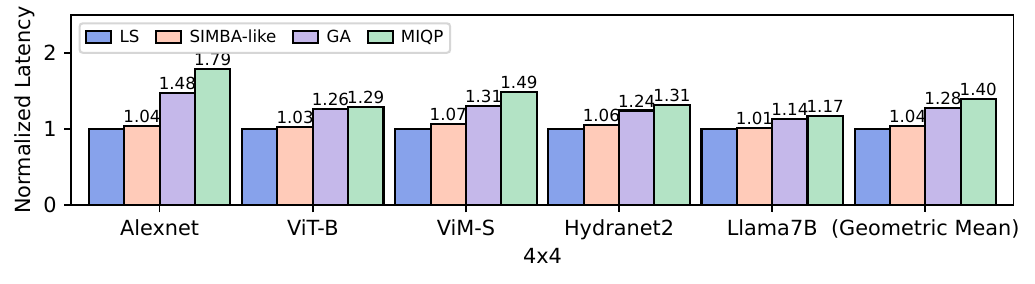}
    \vspace{-1em}
    \caption{Latency}
    \end{subfigure}
    \begin{subfigure}{0.99\linewidth}
    \centering
    \includegraphics[width=0.99\linewidth]{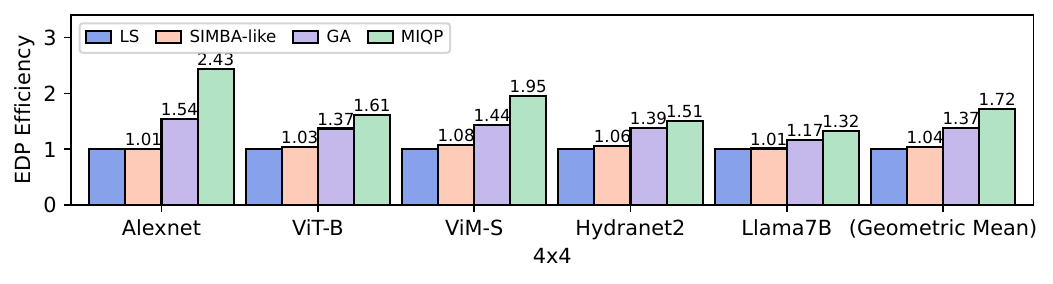}
    \vspace{-1em}
    \caption{EDP}
    \end{subfigure}
    \vspace{-1em}
    \caption{Latency and EDP comparison of MIQP and GA over the baseline for Low-Bandwidth case, 4x4 type-A system. Results are normalized.}
    \label{eval:DRAM}
    \vspace{-1em}
\end{figure}

\autoref{eval:DRAM} demonstrates the performance of GA and MIQP over LS and Simba, with 40\%/28\%, 72\%, and 37\% speedup respectively for latency and edp. In the EDP figure, we can see that the performance gap between GA and MIQP is expanded, compared with the one in \autoref{fig:edp_scaling}. This is because in low-bandwidth cases, part of the congestion transfers to off-chip memory links. Therefore, the complexity brought by on-chip congestion for workload scheduling is reduced, allowing MIQP to find better solutions within the time limit.

\begin{figure}[!t]
    \centering
    \begin{subfigure}{0.99\linewidth}
    \centering
    \includegraphics[width=0.99\linewidth]{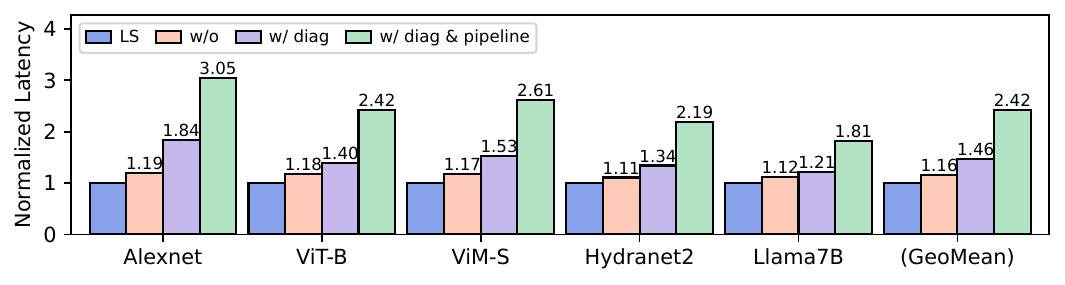}
    \vspace{-1em}
    \caption{Latency}
    \end{subfigure}
    \begin{subfigure}{0.99\linewidth}
    \centering
    \includegraphics[width=0.99\linewidth]{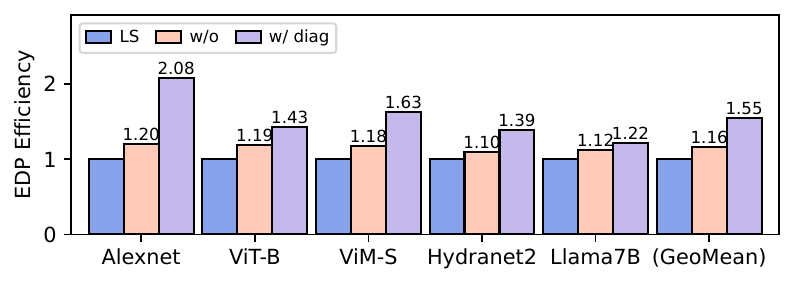}
    \vspace{-1em}
    \caption{EDP}
    \end{subfigure}
    \vspace{-1em}
    \caption{Ablation study on diagonal links and pipelining.}
    \label{eval:ablation}
    \vspace{-1.5em}
\end{figure}

\subsection{Ablation Study}
In \autoref{eval:ablation}, it can be observed that for both latency and EDP tasks, simply doing workload partitioning without diagonal links achieves a relatively small speedup as it cannot bypass congestion during data collection and the latency distribution of access to main memory is more unbalanced, limiting utilization of chiplet far away from global chiplets. Diagonal links greatly alleviate such problems by adding bandwidth to congested places and providing faster access for chiplets originally with large memory latency.

For latency, in addition to the optimized workload partitions, pipelining further utilized idle resources for executions from other samples. Because there exist a lot of locally sequentially chained operators, making use of idle resources significantly reduces the overall latency of the whole batch.

\section{Conclusion}

In this paper, we addressed the challenges of optimizing end-to-end communication and workload partitioning in MCM accelerators.
We proposed a cycle-accurate, congestion-aware, and packaging-adaptive framework.
We propose three optimizations to optimize the framework---diagonal Link, on-package redistribution, and fine-grained pipelining.
We also propose mixed integer linear programming and genetic algorithm based schedulers to solve the optimized framework.
Results show that MCMComm achieves significant EdP~(Energy delay Product) improvement for CNNs and Vision Transformers up to 1.58$\times$ and 2.7$\times$ using GA and MIQP, respectively.

\bibliographystyle{ACM-Reference-Format}
\bibliography{references}

\end{document}